\documentstyle[multicol,amssymb,epsf,prl,aps]{revtex}
\begin{document}
\draft
\title{Shell-model half-lives for the $N=82$ nuclei and their
implications for the r-process}

\author{G. Mart\'{\i}nez-Pinedo and K. Langanke}

\address{Institut for Fysik og Astronomi, {\AA}rhus Universitet,\\
  DK-8000 {\AA}rhus C, Denmark}

\date{\today}
\maketitle

\begin{abstract}
  We have performed large-scale shell-model calculations of the
  half-lives and neutron-branching probabilties of the r-process
  waiting point nuclei at the magic neutron number $N=82$. We find
  good agreement with the measured half-lives of $^{129}$Ag and
  $^{130}$Cd. Our shell-model half-lives are noticeably shorter than
  those currently adopted in r-process simulations. Our calculation
  suggests that $^{130}$Cd is not produced in beta-flow equilibrium
  with the other $N=82$ isotones on the r-process path.
\end{abstract}

\begin{multicols}{2}
  
About half of the elements heavier than mass number $A=60$ are made
in the astrophysical r-process, a sequence of neutron capture and beta
decay processes~\cite{BBFH,Cameron,Thielemann91}. The r-process is
associated with environment of relatively high temperatures ($T
\approx 10^9$ K) and very high neutron densities ($> 10^{20}$
neutrons/cm$^3$) such that the intervals between neutron captures are
generally much smaller than the $\beta$ lifetimes, i.e. $\tau_n \ll
\tau_\beta$ in the r-process. Thus, nuclei are quickly transmuted into
neutron-richer isotopes, decreasing the neutron separation energy
$S_n$. This series of successive neutron captures comes to a stop when
the $(n,\gamma)$ capture rate for an isotope equals the rate of the
destructive $(\gamma,n)$ photodisintegration rate. Then the r-process
has to wait for the most neutron-rich nuclei to $\beta$-decay. Under
the typical conditions expected for the r-process, the $(n,\gamma)
\rightleftarrows (\gamma,n)$ equilibrium is achieved at neutron
separation energies, $S_n \approx 2$ MeV \cite{Thielemann93}.  This
condition mainly determines the r-process path, which is located about
15-20 units away from the valley of stability.  The r-process path
reaches the neutron shell closures at $N=50,82$, and 126 at such low
$Z$-values that $S_n$ is too small to allow the formation of still
more neutron-rich isotopes; the isotopes then have to $\beta$-decay.
To overcome the shell gap at the magic neutron numbers and produce
heavier nuclei, the material has to undergo a series of alternating
$\beta$-decays and neutron captures before it reaches a nucleus close
enough to stability to have $S_n$ large enough to allow for the
continuation of the sequence of neutron capture reactions. Noting that
the $\beta$-decay half-lives are relatively long at the magic neutron
numbers, the r-process network waits long enough at these neutron
numbers to build up abundance peaks related to the mass numbers
$A\approx 80,130$, and 195.  Furthermore the duration of the
r-process, i.e. the minimal time required to transmute, at one site,
seed nuclei into nuclei around $A\approx 200$, is dominantly given by
the sum of the half-lives of the r-process nuclei at the three magic
neutron numbers. It appears as if the required minimal time is longer
than the duration of the favorable r-process conditions in the
neutrino-driven wind from type II supernovae~\cite{woosley}, which is
the currently most favored r-process site.

Simulations of the r-process require a knowledge of nuclear properties
far from the valley of stability. As the relevant nuclei are not
experimentally accessible, theoretical predictions for the relevant
quantities (i.e. neutron separation energies and half-lives) are
needed.  This Letter is concerned with the calculation of
$\beta$-decays of r-process nuclei at the magic neutron number $N=82$.
These $\beta$-decays are determined by the weak low-energy tails of
the Gamow-Teller strength distribution, mediated by the operator
$\bbox{\sigma \tau_-}$, and provide quite a challenge to theoretical
modelling as they are not constrained by sumrules.  Previous estimates
have been based on semi-empirical global models, quasiparticle random
phase approximation, or very recently, the Hartree-Fock-Bogoliubov
method. But the method of choice to calculate Gamow-Teller transitions
is the interacting nuclear shell model, and decisive progress in
programming and hardware make now reliable shell model calculations of
the half-lives of the $N=82$ r-process waiting point nuclei feasible.

Our shell model calculations have been performed with the code {\sc
  antoine} developed by E.~Caurier~\cite{Caurier}. As model space we
chose the $0g_{7/2},1d_{3/2,5/2},2s_{1/2},0h_{11/2}$ orbitals outside
the $N=50$ core for neutrons, thus assuming a closed $N=82$ shell
configuration in the parent nucleus. For protons our model space was
spanned by the $1p_{1/2},0g_{9/2,7/2},1d_{3/2,5/2},2s_{1/2}$ orbitals
where a maximum of 2 (3) protons were allowed to be excited from the
$1p_{1/2},0g_{9/2}$ orbitals to the rest of the orbitals in the parent
(daughter) nucleus. The $0g_{9/2}$ neutron orbit and the $0h_{11/2}$
proton orbit has been excluded from our model space to remove spurious
center of mass configurations. Therefore, we do not allow for
Gamow-Teller transitions from the $0g_{9/2}$ and $h_{11/2}$ neutron
orbitals, which should, however, not contribute significantly to the
low-energy decays we are interested in here. The $1p_{1/2}$ proton
orbit has been included to describe the $1/2^-$ isomeric state seen in
$^{131}$In and expected in the others $N=82$ odd-A isotones, but does
not play any role in the decay of the ground states.  The residual
interaction can be split into a monopole part and a renormalized
G-matrix component which can be derived from the nucleon-nucleon
potential. We use the interaction of ref.~\cite{morten} for the
$gdsh_{11/2}$ orbits and the KLS interaction~\cite{KLS} for the
interaction between the previous orbits with the $1p_{1/2}$ orbit. To
derive the appropriate monopole part, we followed the prescription
given by Zuker~\cite{Zuker}, and fine-tuned the monopoles to reproduce
known spectra of nuclei around the $N=82$ shell closure. As shell
model studies overestimate the GT strength by a universal factor, we
have scaled our results by the appropriate factor
$(0.74)^2$~\cite{quenching}.

The $Q_\beta$ values have been taken either from experiment
($^{131}$In) or from the mass compilation of Duflo and
Zuker~\cite{Duflo}.  Note that the Extended Thomas-Fermi with
Strutinsky Integral approach (ETFSI) \cite{ETFSI} and the
microscopic-macroscopic (FRDM) model of M\"oller \cite{Moeller} give
very similar $Q_\beta$ values (with typical uncertainties of 250 keV
for $Q_\beta\approx 10$~MeV) so that the associated uncertainty in the
half-lives is small.

\begin{figure}
  \begin{center}
    \leavevmode
    \epsfxsize=0.9\columnwidth
    \epsffile{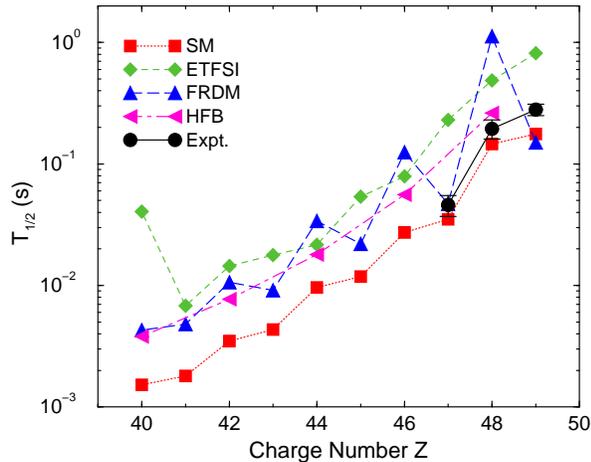} 
    \caption{Comparison of half-lives of the $N=82$ isotones as
      calculated in the FRDM, HFB, ETFSI and the present shell model
      approaches with data.}
    \label{fig1}
  \end{center}
\end{figure}

The shell-model half-lives are summarized in Table~\ref{tab:lives} and
are compared to other theoretical predictions in Fig.~\ref{fig1}. For
$Z=47$-49, the half-lives are known experimentally
\cite{kcd130,Kratz2} and our shell model values are slightly faster.
This, however, is expected as our still truncated model space will
miss some correlations and hence slightly overestimates the
Gamow-Teller matrix elements.

Our shell model half-lives show significant and important differences
to those calculated in the FRDM \cite{Moeller} and the ETFSI
approach~\cite{ETFSI}, which have been typically used in r-process
simulations.  Although the latter predicts a $Z$-dependence of the
half-lives in the $N=82$ isotones very similar to the present results,
the ETFSI half-lives are longer on average by factors 4-5 indicating
that the method fails to shift enough Gamow-Teller (GT) strength to
low energies \cite{ETFSI}.  The FRDM half-lives show a very pronounced
odd-even dependence which is predicted neither by ETFSI nor shell
model.  While the FRDM half-lives for odd-A $N=82$ isotones
approximately agree with the shell model results (within a factor of
2) and the experimental values for $^{131}$In and $^{129}$Ag, they
overestimate the half-lives for even isotones by an order of
magnitude. As such an odd-even dependence is not present in the
experimental half-lives (nor in the r-process abundances) it is
probably an artifact of the FRDM model. The absence of odd-even
effects can be understood considering that the main contribution to
the half-life comes from transitions from a $g_{7/2}$ neutron to a
$g_{9/2}$ proton due to the energy gap between the $g_{9/2}$ and the
other orbits. Therefore, neither the GT matrix elements nor the
half-lifes show a strong odd-even dependence along the $N=82$ isotonic
line. Noting that the $Q_\beta$ values in the FRDM model are very
similar to the ones used and that the main difference with our results
appears when the final nuclei is odd-odd we conclude that the odd-even
effect must stem from the treatment of the $pn$ interaction in the
FRDM approach. Very recently, Engel {\em et al.}  have performed
half-life calculations of r-process nuclei within the HFB
model~\cite{Engel}.  Unfortunately their studies are yet restricted to
even-even nuclei only, but they obtain results which, except for a
factor of 2, closely resemble the present shell model results.  Ref.
\cite{Engel} points out that the half-lives of the $N=82$ waiting
point nuclei are noticeably shorter than currently assumed in
r-process simulations, in support of our findings.

Odd-A nuclei in this mass range usually exhibit a low-lying $1/2^-$
isomeric state which can be related to a proton hole in the $p_{1/2}$
orbital. These isomeric states can affect the r-process half-lives in
two different ways: i) If low enough in energy, the isomeric state can
be populated thermally; ii) in a non-equilibrium picture the isomeric
state can be fed by the preceding neutron capture on the $N=81$
nucleus. The half-live of the isomeric state has been measured in
$^{131}$In (350 ms), very similar to the ground state half-life (280
ms).  We have calculated the energy positions and half-lives of the
isomeric states within our shell model approach. We find that the
excitation energy of the isomeric state slowly decreases within the
$N=82$ isotones when moving from $^{123}$Nb ($E^*=500$ keV) to
$^{131}$In (375 keV) where experimentally only the isomeric state in
$^{131}$In is known (at 360 keV). Importantly our calculation predicts
the half-lives of the isomeric states to be comparable to the ground
state half-lives in all cases (see Table~\ref{tab:lives}). Thus, the
effective r-process half-lives will be very close to the ground state
half-lives. We note that the isomeric state in $^{131}$In dominantly
decays by first-forbidden transitions, as the approximately closed
$g_{9/2}$ proton configuration in this state strongly suppresses
low-energy GT transitions.  We calculate a half-live of the isomeric
state of 274 ms due to first-forbidden decay, about $30\%$ faster than
the experimental value (350 ms). For the $N=82$ nuclei with $Z\le 47$
the $g_{9/2}$ orbital is not anymore closed for the isomeric state
allowing for GT transitions at low-energy. Consequently these nuclei
decay by GT transitions rather than first-forbidden ones.

\begin{figure}
  \begin{center}
    \leavevmode
    \epsfxsize=0.9\columnwidth
    \epsffile{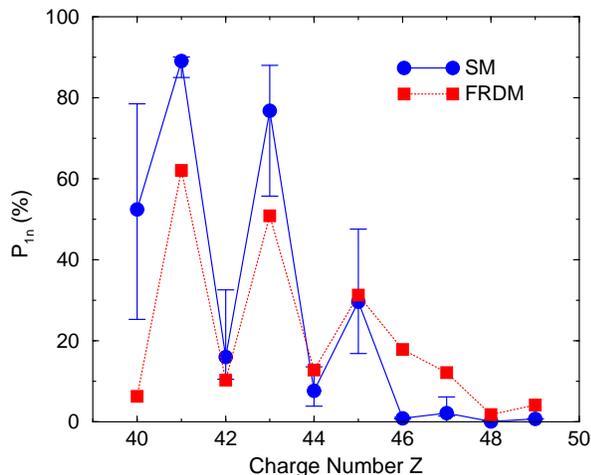} 
    \caption{Comparison of neutron emission probabilities $P_{1n}$ as
      calculated in the FRDM and the present shell model approach.}
    \label{fig2}
  \end{center}
\end{figure}

An interesting, but yet open question is whether the r-process
proceeds in $\beta$-flow equilibrium also at the waiting points
related to magic neutron numbers~\cite{Kratz1}. If so, the duration of
the r-process has to be larger than the sum of the beta half-lives of
the nuclei in $\beta$-flow equilibrium. In this appealing picture, the
observed r-process abundances scale like the respective $\beta$-decay
half-lives, if the former are corrected for $\beta$-delayed neutron
emissions during their decays from the r-process paths towards the
stable nuclei which are observed as r-process abundances. Using our
shell model $\beta$-strength functions we have calculated the
probability $P_{1n}$ that the $\beta$ decay is accompanied by the
emission of a neutron, defined as the relative probability of the
$\beta$-decay rate above the neutron emission threshold, $S_n$.  For
consistency we adopted the $S_n$ values from Duflo and Zuker (DZ),
which for a mild parity effect, gives similar results than the ETFSI
model, while the FRDM model predicts a significantly slower decrease
of $S_n$ with decreasing $Z$. As the $P_{1n}$ values are rather
sensitive to the neutron separation energies we have accounted for the
differences between the ETFSI and the DZ predictions by assigning an
uncertainty of 500 keV to the DZ $S_n$ values. In this way we have
calculated an equally probable range for the shell model $P_{1n}$
values as shown in Fig.~\ref{fig2}. We find rather small neutron
emission probabilities for $Z=46$-49 (which will only slightly been
increased if the $1h$-proton orbitals were included); for the smaller
$Z$-values our results approximately resembles the FRDM values, also
showing a noticeable odd-even dependence.

\begin{figure}
  \begin{center}
    \leavevmode
    \epsfxsize=0.9\columnwidth
    \epsffile{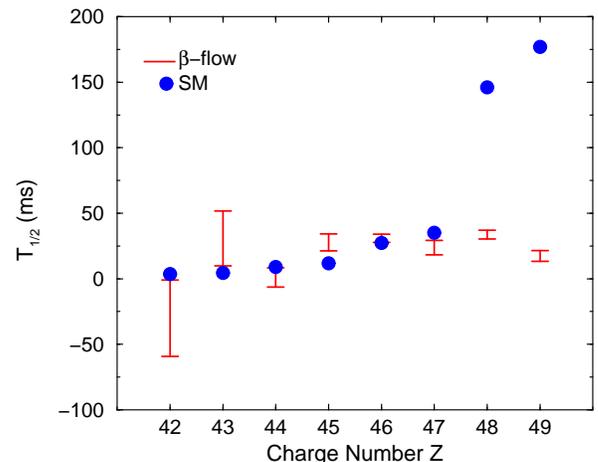} 
    \caption{Test of $\beta$-flow equilibrium by comparing the 
      shell model half-lives for the $N=82$ r-process waiting-point
      nuclei with the $\beta$-flow half-lives. The error bars in the
      latter reflect the errors in the $P_{1n}$ and in the solar
      r-process abundances.}
    \label{fig3}
  \end{center}
\end{figure}

Using the solar r-process abundances~\cite{beer} and the shell model
neutron emission probabilities $P_{1n}$, we have determined the
abundances of the $N=82$ progenitor nuclei, $n(Z)$, on the r-process
path by one-step iteration; we have checked that second-order
branchings change the results only unsignificantly.  If $\beta$-flow
equilibrium at the waiting points is indeed achieved, one has $n(Z)
\sim \tau(Z)$. Thus, up to a constant $n(Z)$ can be expressed as the
so-called $\beta$-flow half-life $T_{\beta f}$. Fixing the constant
appropriately, figure~\ref{fig3} shows that $\beta$-flow equilibrium
can be attained for the $N=82$ isotones with $Z=44$-47, but fails by
more than a factor of 3 for $^{130}$Cd and $^{131}$In. As the
systematic of neutron separation energies put $^{130}$Cd on the
r-process path, our results suggest that the conditions which allow to
build the $N=82$ r-process abundance peak do not last long enough to
achieve $\beta$-flow equilibrium for this nucleus.  This is consistent
with the expectation that the r-process peaks at $N=82$ and $N=126$
are made under different
conditions~\cite{Thielemann93,wasserburg,Qian2}.  Recent observations
also indicate that the nuclides in the $N=82$ and 126 abundance peaks
are produced at different sites~\cite{Cowan}.  Furthermore, the
assumption of $\beta$-flow equilibrium, which on the r-process path
should be fulfilled the better the shorter the half-life, leads to
unphysical (negative) $T_{\beta f}$ values for $Z<43$, indicating that
these nuclei are not on the r-process path.  However, firm conclusion
here can only be reached after reducing the uncertainties in the $P_{1n}$
values for all nuclei in the decay sequence to stability.

So far we have discussed the r-process as a sequence of competing
neutron capture and $\beta$-decay processes. If the r-process site is
indeed the neutrino-driven wind above a newly born neutron star, then
it occurs in a very strong neutrino flux and charged-current
$(\nu_e,e)$ reactions can substitute for $\beta$ decays; in this case
the picture of $\beta$-flow equilibrium has to be extended to
`weak-flow' equilibrium. McLaughlin and Fuller~\cite{Fuller} have
shown that weak steady flow equilibrium at the $N=82$ waiting point
normally cannot be attained in the neutrino-driven wind model.
However, their study adopted $\beta$ half-lives taken from
\cite{Kratz1} which predict a $Z$-dependence of the half-lives in
disagreement with the recent theoretical studies (including the
present) and the data. Nevertheless, when we reinvestigate this
question using the present shell-model $\beta$-decay rates and the
charge current neutrino rates of~\cite{Qian,surmenn}, we find that
$^{129}$Ag and $^{130}$Cd cannot be produce in $\beta$-flow
equilibrium within the neutrino-driven wind model in agreement with
the conclusions reached in ref.~\cite{Fuller}.

In conclusion, we have calculated shell model half-lives and neutron
emission probabilities for the $N=82$ waiting point nuclei in the
r-process, finding good agreement with the experimentally known
half-lives for the $Z=47$-49 nuclei. Our half-lives are significantly
shorter than the ETFSI and FRDM half-lives, which are frequently used
in r-process simulations.  Our results indicate that $^{129}$Ag is
produced in $\beta$-flow equilibrium together with the lighter
isotones at the $N=82$ waiting point. R-process simulations usually
include $^{130}$Cd and even $^{131}$In in the r-process path at
freeze-out. If so, they will not be synthesized in $\beta$-flow
equilibrium. That fact together with the shorter half-lives implies a
shorter waiting time at $N=82$. This is quite welcome to remove
possible conflicts between the required duration time for the
r-process and the expansion time scale of the neutrino-driven wind
scenario which requires the r-process to occur in a fraction of a
second.

We thank E. Caurier, J. Engel, F. Nowacki, F.-K. Thielemann, P. Vogel
and A.\,P. Zuker for useful discussions. This work was supported in
part by the Danish research Council. Computational cycles where
provided by the Supercomputing Facility at the University of {\AA}rhus
and by the Center for Advanced Computational Research at Caltech.

\end{multicols}

\begin{multicols}{2}

\addvspace{-6mm}

\begin{table}
  \caption{Comparison of the shell model half-lives for the ground and 
    isomeric state  with experiment~\protect\cite{Kratz1,Kratz2}. All
    half-lives are in ms.}  
  \label{tab:lives}
  \renewcommand{\arraystretch}{1.1}
  \begin{tabular}{ccccc}
    Nucleus &  \multicolumn{2}{c}{Ground state} &
    \multicolumn{2}{c}{Isomeric state} \\ \cline{2-3} \cline{4-5}
    & Expt.  & Theor. & Expt. & Theor.  \\ \hline
    $^{131}$In  & $280 \pm 30$ & 177 & $350 \pm 50$ & 274 \\
    $^{130}$Cd  & $195 \pm 35$ & 146 & &  \\
    $^{129}$Ag  & $46 \pm 9$ & 35.1 & & 39.5  \\
    $^{128}$Pd  &  & 27.3 & &  \\
    $^{127}$Rh  &  & 11.8 &  & 12.4  \\
    $^{126}$Ru  &  & 9.6  & &  \\
    $^{125}$Tc  &  & 4.3  & & 4.2  \\
    $^{124}$Mo  &  & 3.5  & &  \\
    $^{123}$Nb  &  & 1.8  & & 1.7  \\
    $^{122}$Zr  &  & 1.5  &  \\
  \end{tabular}
\end{table}

\end{multicols}

\end{document}